\documentclass[fleqn,usenatbib]{mnras}

\usepackage{newtxtext,newtxmath}

\usepackage[T1]{fontenc}

\DeclareRobustCommand{\VAN}[3]{#2}
\let\VANthebibliography\thebibliography
\def\thebibliography{\DeclareRobustCommand{\VAN}[3]{##3}\VANthebibliography}

\usepackage{threeparttable}
\usepackage{xspace}
\usepackage{bm}
\usepackage{commath}

\usepackage{graphicx}	
\usepackage{amsmath}	

\newcommand{\mdet}{\textsc{metadetection}\@\xspace}
\newcommand{\mcal}{\textsc{metacalibration}\@\xspace}

\newcommand{\dmcal}{\textsc{deep-field~metacalibration}\@\xspace}

\newcommand{\fpfs}{\texttt{FPFS}\@\xspace}
\newcommand{\acal}{\texttt{AnaCal}\@\xspace}
\newcommand{\dacal}{\textsc{deep-field~}\texttt{AnaCal}\@\xspace}
\newcommand{\dtilde}[1]{\tilde{\tilde{#1}}}

\title[Deep-Field Analytical Calibration]{Deep-Field Analytical Calibration}

\author[A. Park et al.]{
Andy Park,$^{1, 2}$\thanks{E-mail: chanhyup@andrew.cmu.edu}
Xiangchong Li,$^{3}$
Rachel Mandelbaum$^{1}$
and Matthew Becker$^{2}$
\\
$^{1}$McWilliams Center for Cosmology and Astrophysics, Department of Physics, Carnegie Mellon University, 5000 Forbes Ave, Pittsburgh, PA 15213\\
$^{2}$High Energy Physics Division, Argonne National Laboratory, Lemont, IL 60439, USA\\
$^{3}$Brookhaven National Laboratory, Bldg 510, Upton, New York 11973, USA
}

\date{Accepted XXX. Received YYY; in original form ZZZ}

\pubyear{\the\year{}}

\begin{document}
\label{firstpage}
\pagerange{\pageref{firstpage}--\pageref{lastpage}}
\maketitle

\begin{abstract}
    The next generation of imaging surveys, including the Vera C. Rubin Observatory Legacy Survey of Space and Time (LSST), Euclid, and the Nancy Grace Roman Space Telescope, will provide unprecedented constraints on cosmology using weak gravitational lensing. To fully exploit the statistical power of these surveys, shear measurement methods must achieve sub-percent accuracy while mitigating a range of systematic biases, including those arising from noise, the point-spread function (PSF), blending, and shear-dependent detection effects. The latest version of the analytical calibration framework (\acal) has been demonstrated to achieve such accuracy, but requires the addition of noise to the images, reducing their effective depth. 
    In this work, we introduce Deep-Field Analytical Calibration (\dacal), an extension of \acal that exploits deep-field images to compute shear responses while preserving the statistical power of wide-field images. We validate \dacal on both isolated and blended galaxy image simulations with LSST-like seeing and noise levels,
showing that it meets the stringent requirement of a multiplicative bias $|m| < 3\times10^{-3}$ at the 99.7 percent confidence level. Compared to standard \acal applied on wide-field images, \dacal improves the effective galaxy number density from $17$ to $30$ galaxies per arcmin$^2$ for simulated 10-year LSST images. 
If the deep field has $10\times$ the exposure time of the wide field,  
we find the pixel noise variance in the shear estimation is reduced by $30\%$ and the uncertainty in shear estimation is reduced by approximately $25\%$.  
Finally, we assess the impact of sample variance on \dacal, following the LSST Deep Drilling Fields observing strategy, and find that its equivalent calibration uncertainty is $\lesssim 0.3\%$. Our results show that \dacal is a promising approach for shear calibration in upcoming weak lensing surveys. 
\end{abstract}




\section{Introduction}
\label{sec:introduction}
The next generation of imaging surveys, such as the Vera C.\ Rubin Observatory Legacy Survey of Space and Time \citep[LSST;][]{LSST_ivezic2019}, \textit{Euclid} \citep{Euclid_Laureijs}, and \textit{Nancy Grace Roman Space Telescope} High Latitude Imaging Survey \citep{Roman_Akeson}, will provide unprecedented constraints on the growth of large scale structure and the fundamental properties of the Universe. A key observable in these surveys is weak gravitational lensing, or cosmic shear, which probes the distribution of matter by measuring the coherent distortions in the shapes of distant galaxies induced by intervening large-scale structure (see \citealt{cosmicshear_Kilbinger2015, precisioncosmology_mandelbaum} for a review of cosmic shear). To fully exploit the statistical power of these surveys, shear measurement techniques must achieve sub-percent accuracy while mitigating a range of systematic biases, including those arising from the point-spread function (PSF), noise biases, model biases, selection effects, blending effects, and detection biases (see \citealt{precisioncosmology_mandelbaum} for a review of systematics in weak lensing and references).

Several advanced shear estimation methods have been developed to reduce the reliance on external simulations in calibration. These methods include \texttt{BFD} \citep{BFD_Berinstein2016}, \mcal \citep{Metacal2017_sheldon, Metacal_HuffMandelbaum}, and \fpfs \citep{fpfs_1, fpfs_2}. These methods have demonstrated the ability to reach the required sub-percent accuracy for isolated galaxies. \citealt{metaDet_LSST2023} extended \mcal and introduced the state-of-the-art technique \mdet and \citealt{fpfs_3, fpfs_4} further extended \fpfs to include an analytical calibration technique and introduced \acal. Both of these newly developed shear estimators are able to reach the sub-percent accuracy in the presence of blending.
Because it is an analytical approach, \acal is more than a hundred times faster than the \mdet algorithm.

The previous version of \fpfs \citep{fpfs_3} corrected the noise bias by computing the second-order derivative (Hessian matrix) of the ellipticity. However, this approach required estimating noisy second- and higher-order derivatives, introducing additional complications from higher-order noise terms. The current implementation of \mdet and \acal achieves the required accuracy in shear measurement by introducing an additional layer of noise, matched to the statistical properties of the original image, to cancel out noise bias. This process effectively doubles the variance of pixel noise in survey images ($\sqrt{2}$ in the noise standard deviation) or equivalently reduces the signal-to-noise ratio of every source by approximately $30\%$.

Modern imaging surveys typically allocate the majority of their observing time, over 90 percent, to wide-field imaging, while dedicating a smaller fraction to deep-field imaging, which covers a limited area, but receives a significantly higher number of visits per observing night. This strategy was used in the Dark Energy Survey deep-fields \citep{DES_deep} and is planned for the LSST Deep Drilling Fields  \citep[DDFs;][]{LSST_deep} and Euclid Deep Fields \citep{Euclid_deep}. 
Wide-field surveys provide the statistical constraining power necessary for cosmic shear studies but are limited by the higher level of noise in the images.
In contrast, deep-field observations capture fainter galaxies, improve photometric redshifts, and enable more precise shape measurements due to lower noise. 

Both the \mdet and \acal ensemble shear estimation methods consist of two key components: the uncalibrated shape measurement, $\langle\bm{e}\rangle$, and the mean shear response matrix, $\langle\bm{R}\rangle$. The mean shear estimate is then obtained as \citep{Metacal_HuffMandelbaum, Metacal2017_sheldon, fpfs_3}
\begin{equation}
    \hat{\gamma} \equiv \frac{\langle \bm{e}\rangle}{\langle\bm{R}\rangle}.
\end{equation}
Traditional analyses derive both components using wide-field survey data, where the dominant source of uncertainty 
in shear estimation arises from the shape measurement noise. \citet{deep_metacal} introduced \dmcal, a novel technique in which the mean response matrix $\langle\bm{R}\rangle$ is computed using deep-field survey data instead of the wide-field survey data. This approach effectively confines the doubling in the pixel noise variance to the deep-field survey, avoiding additional noise contributions in the higher-noise wide-field data. They demonstrated that \dmcal yields an unbiased shear estimate while reducing the pixel noise variance in images by $\sim$30\%, increasing the signal-to-noise of lensing sources and reducing pixel noise contributions to the uncertainty in shear estimation. 
However, the technique was only demonstrated on isolated galaxies. Building upon this approach, we introduce \dacal, an extension of \acal, that calibrates the shape measurement using the shear responses derived from the deep-field survey data. Furthermore, we generalize this technique to account for the effects of blending and incorporate corrections for shear-dependent detection bias.

This paper is organized as follows. In Section~\ref{sec:dacal} we give a brief overview of the \fpfs shear estimator within the \acal framework and introduce the formalism for \dacal. 
In Section~\ref{sec:result:simulaiton} we present galaxy image simulations that we used to test our method. In Section~\ref{sec:result}, we cover the main results of our work and tests of shear calibration for isolated galaxies (Sec.~\ref{sec:result:isolated}) and blended galaxies (Sec.~\ref{sec:result:blended}). In Section~\ref{sec:result:sample_variance}, we address the issue of sample variance in the deep-field images. Finally, in Section~\ref{sec:conclusion}, we summarize our results and future outlook.

\section{\acal Formalism}
\label{sec:dacal}


In this section, we first briefly introduce the standard \fpfs shear estimator developed in \citet{fpfs_1, fpfs_2} and corrected for detection bias \citep{fpfs_3, fpfs_4} and noise bias \citep{anacal_li} within the \acal framework. We then introduce the formalism defining \dacal.

In this paper, we simplify the notation and use reduced shear and shear indistinguishably by setting the lensing convergence to zero. Shear is denoted by a complex spinor $\gamma = \gamma_1 + \mathrm{i}\gamma_2$, where $\mathrm{i}$ is the imaginary unit, $\gamma_1$ is the amount of stretching of the image along the horizontal / vertical direction, and $\gamma_2$ is the amount of stretching of the image in the direction at an angle of $45 \deg$ with the horizontal direction. In the weak lensing regime, the shear magnitude is typically on the order of a few percent or less ($|\gamma| \lesssim 0.02$) \citep{Shapiro_2009}. The shear signal is much smaller than the shape noise arising from the intrinsic shapes of galaxies, requiring a large ensemble of galaxies to accurately infer shear \citep{precisioncosmology_mandelbaum, Prat_2025review}. 

\subsection{Standard \fpfs within \acal Framework}
\label{sec:dacal:fpfs}
We adopt the approach outlined in \cite{fpfs_3} to define galaxy shape, size, flux, and the detection process using a set of linear observables. These observables are constructed as linear combinations of pixel values and include both shapelet modes \citep{shapelets_modes_Refregier2003, polar_shapelets_Massey2005} and detection modes \citep{fpfs_3}. We define the vector $\bm{\nu} = (\nu_1, \nu_2, \dots, \nu_n)$ to represent the set of linear observables. For a galaxy profile $f(\bm{x})$, with $\bm{x}$ denoting coordinates in real space, the linear observables are the projection of image onto basis kernels after PSF deconvolution in Fourier space. For each linear observable, we define

\begin{equation}
\label{eq:linear_observable}
    \nu_i = \iint \mathrm{d}^2k \: \frac{f^p(\bm{k})}{p(\bm{k})} \chi_i(\bm{k}),
\end{equation}
where $f^p(\bm{k})$ is the observed (PSF-convolved, noiseless) image in Fourier space, $p(\bm{k})$ is the PSF image in Fourier space, and $\chi_i(\bm{k})$ are the basis kernels.

Galaxy detection in \fpfs is based on four detection modes ($\nu_i$ with $i = 0, 1, 2, 3$) per pixel. These detection modes characterize differences in the values of pixels relative to the nearby pixels in four orthogonal directions. These detection modes are then used to identify peaks. The selection bias introduced by this detection process is analytically corrected using the shear responses of the pixel values associated with these peak modes following \cite{fpfs_3}. The basis kernel for detection, $\psi_i(\bm{k})$ for wave number vector $\bm{k} = (k_1, k_2)$, is defined in Fourier space as
\begin{equation}
    \label{eq:detection_modes}
    \psi_i = \frac{1}{(2\pi)^2} e^{-(k_1^2+k_2^2) \sigma_h^2/2} \left(1 - e^{\mathrm{i}(k_1 x_i + k_2 y_i)}\right),
\end{equation}
where $(x_i, y_i) = (\cos (i\pi/2), \sin(i\pi/2))$ and $i \in \{0, 1, 2, 3\}$ are the four nearby pixels in four directions separated by $\pi/2$. $\sigma_h$ is the smoothing scale of shapelets and the detection kernel. Typically, $\sigma_h$ is chosen to be larger than the scale radius of the PSF in the real space. This ensures that the deconvolution process does not amplify noise on small scales or large $|\bm{k}|$ values.

\fpfs uses polar shapelet \citep{polar_shapelets_Massey2005} to derive various galaxy properties, including flux, size, and shape. The basis kernel for source measurement, $\phi_{nm}(\bm{x})$ for position $\bm{x}$ in configuration space, are defined in polar coordinates $(\rho, \theta)$ as
\begin{equation}
\label{eq:shapeletbasis}
\begin{split}
\phi_{nm}(\bm{x}\:|\:\sigma_h) &= (-1)^{(n - |m|)/2}\left\{\frac{[(n-|m|)/2]!}{[(n+|m|)/2]!}\right\}^{\frac{1}{2}}\\
&\times\left(\frac{\rho}{\sigma_h}\right)^{|m|}L_{\frac{n-|m|}{2}}^{|m|}\left(\frac{\rho^2}{\sigma_h^2}\right) e^{-\rho^2/2\sigma_h^2}e^{im\theta},
\end{split}
\end{equation}
where $L_{\frac{n-|m|}{2}}^{|m|}$ are the Laguerre  polynomials, $n$ is the radial quantum number and can be any non-negative integer and $m$ is the spin number which is an integer between $-n$ and $n$ in steps of two. The shapelet modes $M_{nm}$ are obtained by projecting the deconvolved image onto the polar shapelet functions. 
In summary, detection modes are computed by substituting the detection kernel $\psi_i(\bm{k})$ from equation~\eqref{eq:detection_modes} into equation~\eqref{eq:linear_observable}. For source measurements such as flux, size, and shape, the Fourier transform of the shapelet basis $\phi_{nm}(\bm{x} | \sigma_h)$ from equation~\eqref{eq:shapeletbasis} is substituted into equation~\eqref{eq:linear_observable} to obtain the corresponding shapelet modes $M_{nm}$. 

Under the weak lensing limit, the linear observables have a nice property in that they have analytically solvable linear responses to shear, and the $i$th component of the linear shear responses with respect to the shear component $\alpha$ is:
\begin{equation}
    \label{eq:linear_observable_response}
    \nu_{;\alpha i} \equiv \frac{\partial \nu_i}{\partial \gamma_\alpha}
    = \iint \mathrm{d}^2k \frac{f^p(\bm{k})}{p(\bm{k})} \chi_{;\alpha i}(\bm{k})\,.
\end{equation}

The shear response of each shapelet basis can be written as a linear combination of shapelet basis given in \cite{polar_shapelets_Massey2005} and detection kernel are given in \cite{fpfs_3}.

The linear observables $\nu$ are then passed into nonlinear functions to define nonlinear observables $e$:
\begin{equation}
\label{eq:nonlinear_ellip}
    e(\bm{\nu}) = \epsilon(\bm{\nu})w_s(\bm{\nu})w_d(\bm{\nu}),
\end{equation}
where $\epsilon$ is the spin-2 \fpfs ellipticity spinor defined in, $w_s$ is the selection weight, and $w_d$ is the detection weight. The weight of zero corresponds to rejecting the galaxy from the sample. The selection weight uses shapelet modes to select bright galaxies with a high signal-to-noise ratio and well-resolved large galaxies and the detection weight defines cut on peak modes. Since the shear responses of the detection modes and shapelet modes is analytically obtained, we can analytically compute the derivative of the selection and detection weight function with respect to shear. We adopt the same smoothstep functions of the linear observables and their hyperparameters defined in \cite{anacal_li}.

The spin-2 \fpfs ellipticity is defined as
\begin{equation}
    \epsilon = \epsilon_1 + \mathrm{i}\epsilon_2 \equiv\frac{M_{22}}{M_{00} + C},
\end{equation}
where the weighting parameter $C$ is a constant that adjusts the relative weight for galaxies with different brightness. \citet{FourthOrderShear_Park} extended \citealt{fpfs_3} and used fourth-order shapelet moments $M_{42}$ to define spin-2 \fpfs ellipticity, and combined the shear estimation from the second-order shapelet moment and fourth-order shapelet moment to reduce the galaxy shape noise. In this work, we only consider the ellipticity defined using the second-order shapelet moment.

The linear shear response matrix, $\frac{\partial e_i}{\partial \gamma_j}$, where $i,j\in\{1,2\}$ represent the components of spin-2 quantities, can be derived from the linear shear responses of the linear observables by applying the chain rule:
\begin{equation}
    \frac{\partial e_i}{\partial \gamma_j} = \sum_k \frac{\partial e_i}{\partial \nu_k}\frac{\partial \nu_k}{\partial \gamma_j} = \sum_k \frac{\partial e_i}{\partial \nu_k}\nu_{;jk}.
\end{equation}

Real images have pixel noise, causing measurement error in shear estimation.
\citet{Metacal2017_sheldon} first introduced a numerical recipe to correct for sheared noise by adding an additional layer of noise to the already-noise image, and \citet{anacal_li} analytically proved that the method is free from noise bias. This additional layer of noise has the same noise statistical properties after being rotated counterclockwise by 90 degree, with the rotation defined in the pre-PSF space. We denote the measurement error on the linear observable as $\delta\bm{\nu}$, then we have each measurement error on each linear observable as
\begin{equation}
    \label{eq:measurement_error}
    \delta\nu_i = \iint \mathrm{d}^2k \frac{n(\bm{k})}{p(\bm{k})}\chi_i(\bm{k}),
\end{equation}
where $n(\bm{k})$ is the pure noise image in Fourier space. Adding an additional layer of noise to the image introduces an extra measurement error, $\delta\bm{\nu}'$, on the linear observables, where define the measurement error on each linear observable using equation~\eqref{eq:measurement_error} but with $n'(\bm{k})$ in the integrand instead. Note that $\delta\bm{\nu}$ is the measurement error in actual images and it cannot be directly measured, but $\delta\bm{\nu}'$ can be measured because the pure noise $n'(\bm{k})$ is generated from simulation and we know the exact noise realization. The non-linear ellipticity is then measured on the noisy linear observables $\dtilde{e} = e(\dtilde{\bm{\nu}})$ with $\dtilde{\bm{\nu}} = \bm{\nu} + \delta\bm{\nu} + \delta\bm{\nu}'$. Similar to equation~\eqref{eq:linear_observable_response}, the shear response of renoised linear observables is:
\begin{equation}
    \dtilde{\nu}_{;\alpha i} = \iint \mathrm{d}^2k \frac{f^p(\bm{k}) + n(\bm{k}) + n'(\bm{k})}{p(\bm{k})} \chi_{;\alpha i}(\bm{k}),
\end{equation}
where $i$ is the $i$th component of the linear observable and $\alpha$ is the component of the shear. Since image noise and the additional layer of noise are not affected by shear, the shear responses of $\delta\bm{\nu}$ and $\delta\bm{\nu}'$ are zero. To correct for the noise bias, we can derive:
\begin{equation}
    \frac{\partial \dtilde{\nu}_i}{\partial\gamma_\alpha} = \dtilde{\nu}_{;\alpha i} - \delta\nu_{;\alpha i} - \delta\nu'_{;\alpha i},
\end{equation}
and following the proof from \citet{anacal_li}, the shear response of the renoised ellipticity can be measured after adding the simulated noise as (adopting Einstein notation)
\begin{equation}
    \label{eq:acal_response}
    \left\langle\dtilde{R}_\alpha\right\rangle = \left\langle\frac{\partial e_\alpha(\dtilde{\bm{\nu}})}{\partial\dtilde{\nu}_i}\left(\dtilde{\nu}_{;\alpha i}  - 2\delta\nu'_{;\alpha i}\right)\right\rangle,
\end{equation}
and the shear estimator is
\begin{equation}
    \hat{\gamma}_\alpha = \frac{\langle\dtilde{e}_\alpha\rangle}{\left\langle\dtilde{R}_\alpha\right\rangle} + \mathcal{O}(\gamma^3)\,.
\end{equation}
\citet{anacal_li} showed that this shear estimator is free from noise bias and is accurate to second order of shear. A limitation of this method is that we double the image noise before running detection and source measurement.

\begin{figure*}
    \centering
    \includegraphics[width=\linewidth]{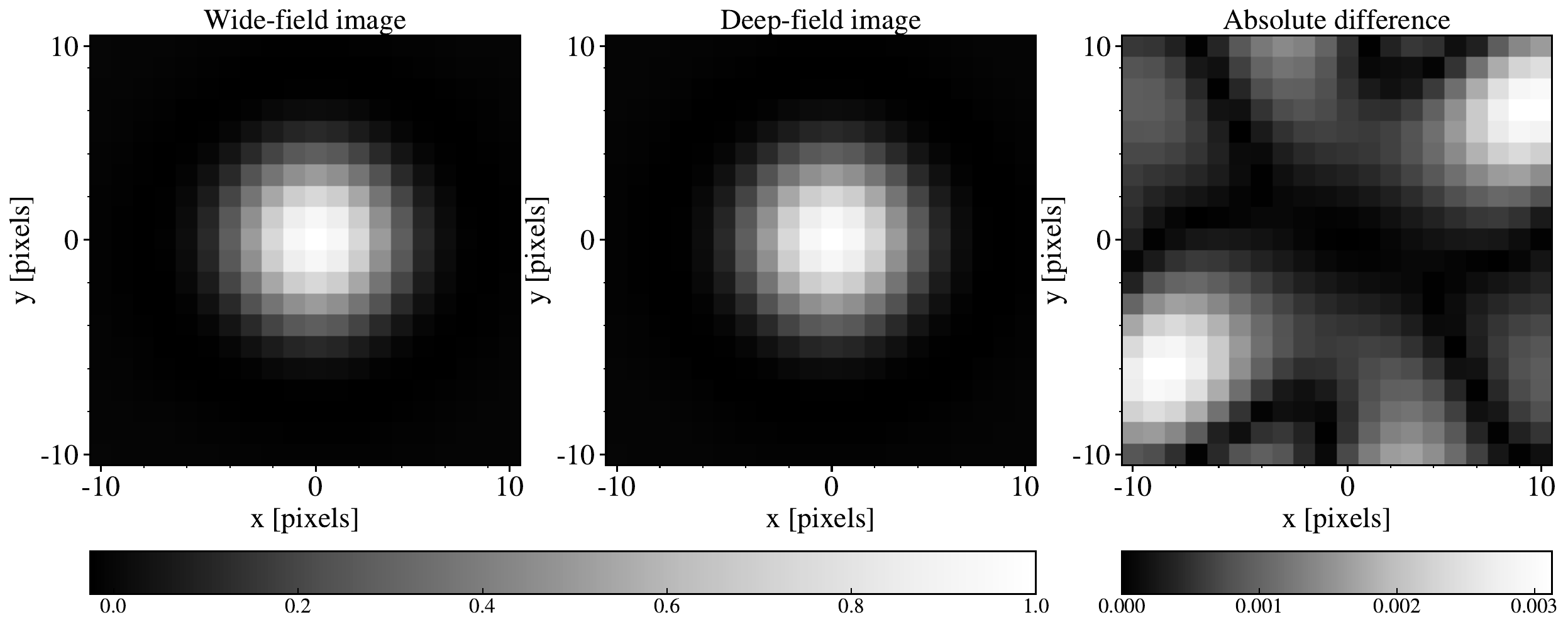}
    \caption{This plot demonstrates that our implementation of \dacal produces consistent noise properties across wide- and deep-field images, as expected from equations~\eqref{eq:nu_wide} and~\eqref{eq:nu_deep}. The left and middle panel show the noise correlation function in configuration space after applying \dacal. The correlation function is normalized such that the maximum is 1. 
    The right panel presents the absolute difference in the noise correlation function computed from $10^5$ deep- and wide-field images. The observed differences are on the order of $10^{-3}$, indicating the good numerical accuracy of our implementation.}
    \label{fig:noise_matching}
\end{figure*}

\subsection{\dacal}
\label{sec:dacal:dacal}
One of the primary goals of \dacal is to avoid doubling the variance in the wide-field images. \dacal achieve this goal by instead doubling the variance in the deep-field images.
As outlined earlier, the shape of the galaxy would be computed from the wide-field images and the shear response would be derived from the deep-field images, which typically has lower image pixel noise variance compared to the wide-field images. To account for common sources of bias in the shear estimates, it is essential to match the PSF and noise distributions between the deep- and wide-field images. A straightforward method to ensure consistency in noise distributions is to apply the wide-field noise to the deep-field images and vice versa. To match the PSF, we deconvolve the galaxy with their respective PSF model and reconvolve with a larger PSF than the both deep- and wide-field PSFs.

\begin{align}
\label{eq:nu_wide}
\nonumber\dtilde{\nu}^{\mathrm{wide}}_i &= \iint \mathrm{d}^2k \left(\frac{f_\mathrm{wide}^p(\bm{k}) + n_\mathrm{wide}(\bm{k})}{p_\mathrm{wide}(\bm{k})} + \frac{n_\mathrm{deep}'(\bm{k})}{p_\mathrm{deep}(\bm{k})} + \frac{n_\mathrm{deep}''(\bm{k})}{p^{90}_\mathrm{deep}(\bm{k})}\right)\chi_i(\bm{k})\\
&= \nu_i^\mathrm{wide} + \delta\nu_i^\mathrm{wide} + \delta\nu_i^{\mathrm{deep},'} + \delta\nu_i^{\mathrm{deep},''}\\
\label{eq:nu_deep}
\nonumber\dtilde{\nu}^\mathrm{deep}_i &= \iint \mathrm{d}^2k \left(\frac{f_\mathrm{deep}^p(\bm{k}) + n_\mathrm{deep}(\bm{k})}{p_\mathrm{deep}(\bm{k})} + \frac{n_\mathrm{deep}'(\bm{k})}{p^{90}_\mathrm{deep}(\bm{k})} +  \frac{n_\mathrm{wide}'(\bm{k})}{p_\mathrm{wide}(\bm{k})}\right)\chi_i(\bm{k})\\
&= \nu_i^\mathrm{deep} + \delta\nu_i^\mathrm{deep} + \delta\nu_i^{\mathrm{deep},'} + \delta\nu_i^{\mathrm{wide},'},
\end{align}
where $f^p_{\mathrm{wide},\mathrm{deep}}(\bm{k})$ are the noiseless PSF-convolved wide/deep-field images in Fourier space, $p_{\mathrm{wide},\mathrm{deep}}(\bm{k})$ are the wide/deep-field PSFs in Fourier space, $p^{90}_{\mathrm{wide}, \mathrm{deep}}$ are the PSF after being rotated by counterclockwise by $90$ degree, $n_{\mathrm{wide},\mathrm{deep}}(\bm{k})$ are the wide/deep-field image noise fields, and $n_{\mathrm{wide},\mathrm{deep}}^{',''}(\bm{k})$ are the pure wide/deep-field noise realizations used for renoising method. The $90$ degree rotation is applied to suppress residual spin-2 anisotropies in the noise image after deconvolution. Note that the scale radius of the kernel $\sigma_h$ is chosen so that it is larger than the size of each wide- and deep-field PSF size. The resulting linear observables of \dacal have a matching PSF and noise distribution with an image noise variance of $\sigma_\mathrm{wide}^2 + 2 \sigma_\mathrm{deep}^2$. The shear response of \dacal then is
\begin{equation} 
\label{eq:dacal_response}
\left\langle\dtilde{R}_{\mathrm{dacal},\alpha}\right\rangle = \left\langle\frac{\partial e_\alpha(\dtilde{\bm{\nu}}^\mathrm{deep})}{\partial \dtilde{\nu}_i^\mathrm{deep}}\left(\dtilde{\nu}^{\mathrm{deep}}_{;\alpha i} - 2\delta\nu^{\mathrm{deep},'}_{;\alpha i} - \delta\nu^{\mathrm{wide},'}_{;\alpha i} \right)\right\rangle
\end{equation}
and the nonlinear observable $\bm{e}$ becomes $\dtilde{\bm{e}}_\mathrm{dacal} = \bm{e}(\dtilde{\nu}^\mathrm{wide})$. If the shear calibration is performed using \acal on just wide-filed images alone, the shear response incorporates two instances of wide-field noise as shown in equation~\eqref{eq:acal_response}. In contrast, \dacal leverages both wide- and deep-field images; the noise contributions on shear response in \dacal consist of one instance of wide-field noise and two instances of deep-field noise. The shear estimator is then
\begin{equation}
    \hat{\gamma}_\alpha = \frac{\langle \dtilde{e}_{\mathrm{dacal}, \alpha}\rangle}{\left\langle \dtilde{R}_{\mathrm{dacal}, \alpha}\right\rangle} + \mathcal{O}(\gamma^3)
\end{equation}

\begin{figure*}
    \centering
    \includegraphics[width=0.8\linewidth]{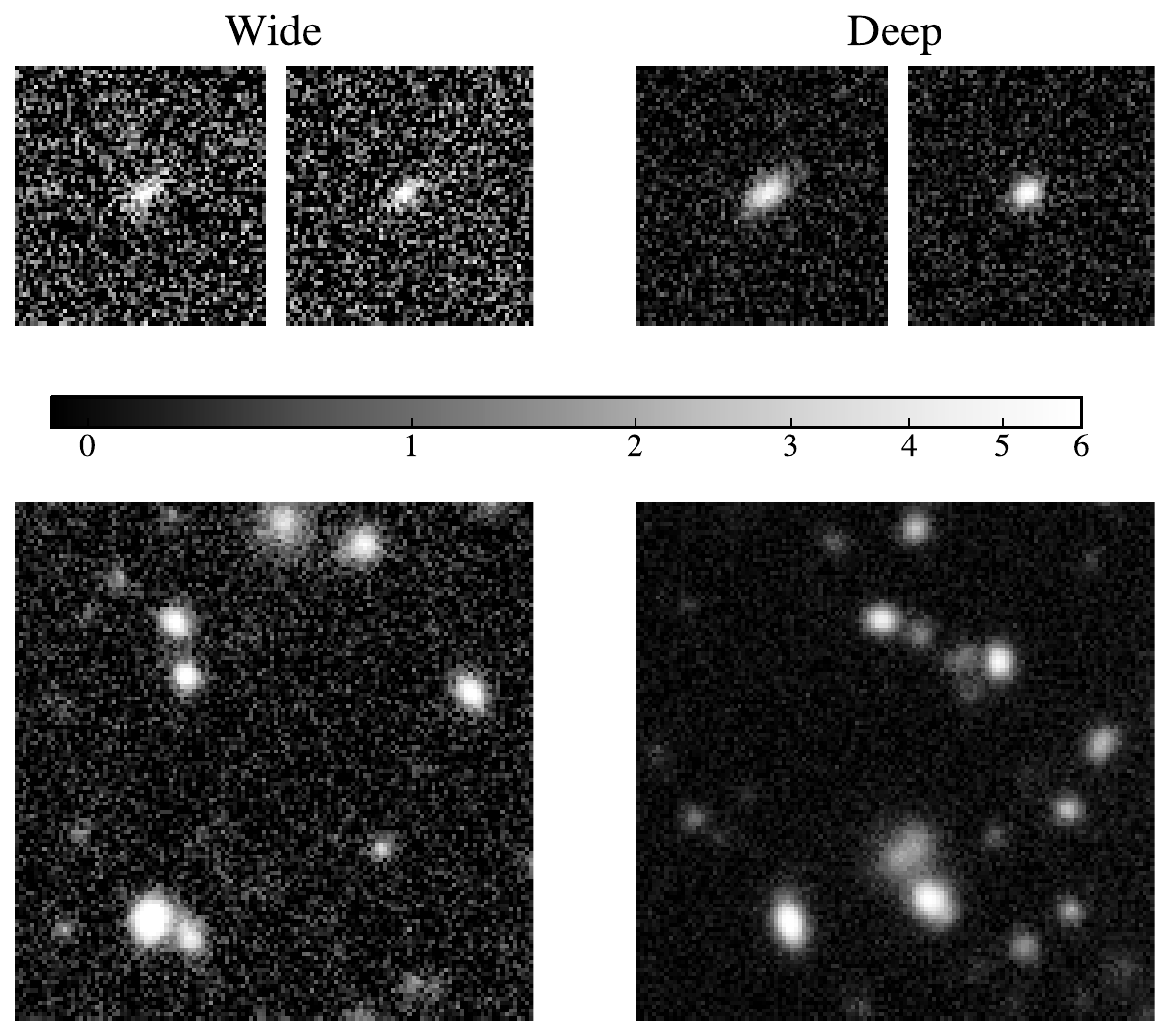}
    \caption{Comparison of wide-field (left) and deep-field (right) images. The top row shows isolated galaxy simulations, while the bottom row presents blended galaxy simulations. All images were simulated with a PSF FWHM of 0.7 arcsec. The isolated images (top) show two $64\times64$ pixel stamps for each field. 
    The blended images (bottom) are randomly located cut-out coadded images of $128 \times 128$ pixels (equivalent to $0.43 \times 0.43$ arcmin$^2$) in the  $r$ band  \citep{metaDet_LSST2023}. The noise variance for wide-field image simulations is set to the mean noise variance in the 10-year LSST observation. 
    The galaxy images in the deep field have SNR that is a factor $\sqrt{10}$ higher than the same galaxy images in the wide field. 
    }
    \label{fig:sample_wide_deep}
\end{figure*}

\section{Simulation setup}
\label{sec:result:simulaiton}
We compute the multiplicative and additive shear biases \citep{mandcbiases_heymans2006} to quantify the accuracy of the shear recovery of \dacal. The estimated shear, $\hat{\gamma}$, is related to the true shear, $\gamma$, as
\begin{equation}
    \label{eq:mcbias}
    \hat{\gamma}_\alpha = (1 + m_\alpha)\gamma_\alpha + c_\alpha,
\end{equation}
where the subscript $\alpha$ denotes the each component of the shear, $m_\alpha$ is the multiplicative bias and $c_\alpha$ is the additive bias. We report errors with $3\sigma$. To reduce shape noise in the estimation of $m$ and $c$ in the image simulation, we adopt the 90-degree rotation \citep{90degreerot_massey2007} and simulated different sheared versions of the same image simulation \citep{preciseSim_Pujol2019} to compute $m_\alpha$ and $c_\alpha$. 

To test the accuracy and precision of \dacal, we generate two sets of image simulations: isolated galaxies, where each realization contains multiple postage stamps with each stamp having one galaxy at the center, and blended galaxies, where galaxies are randomly located. 
The pixel scale for each type of image simulation is set to 0.2 arcsec and we use the \citet{moffat} profile,
\begin{equation}
    p_m(\bm{x}) = \left[1 + c\left(\frac{|\bm{x}|}{r_p}\right)\right]^{-2.5},
\end{equation}
where $\bm{x}$ is the position vector relative to the center of the profile and $c$ and $r_p$ are set such that the full width half maximum (FWHM) of the Moffat profile matches the desired FWHM for the simulation. Fig.~\ref{fig:sample_wide_deep} shows a small region of one simulated image for both wide-field and deep-field simulations using FWHM of 0.7 arcsec. In this work, we set $\sigma_h = 0''.52$ to maximize the effective number density of galaxy according to \cite{fpfs_4}. To save computational time, we only test the accuracy and precision of \dacal using the first component of shear, $\hat{\gamma}_1$, and drop the subscript ($m\equiv m_1$ and $c\equiv c_1$); the multiplicative and additive bias of $\hat{\gamma}_2$ should be comparable to those of $\hat{\gamma}_1$.  

\subsection{Isolated galaxies}

We evaluate the performance of \dacal using isolated galaxy image simulations. This analysis aims to quantify the accuracy of \dacal in the absence of blending and shear-dependent detection effects, and only focus on the impact of shear, PSF size, and image noise. We generate galaxy images using publicly available software \textsc{Galsim}  \citep{galsim} and the COSMOS HST parametric galaxy catalog \citep{mandelbaum_2019_3242143}, adopting a limiting magnitude of $F814W=25.2$ for the input catalog. At this limit, some lower SNR galaxies expected in the isolated simulations may be absent from the analysis.

Galaxies are rendered into $64\times64$ pixel postage stamp images after applying constant shear distortion and PSF convolution. Each simulated image realization (subfield) consists of a $100\times100$ grid of postage stamps. When quantifying the accuracy of the shear estimator, we cancel out the intrinsic shape noise in our simulation by having our galaxy sample contain orthogonal galaxies with the same morphology and brightness, but with the intrinsic major axes rotated by $90 \deg$.

The simulations span four test cases (with results summarized in Table~\ref{tab:iso_setup}) that progressively introduce variability in noise, PSF, and galaxy morphology. In all cases, \dacal is provided with the true PSF model and the noise variance used in the simulations. This setup ensures that any residual calibration bias arises from sensitivity to mismatches in image properties, not from misestimate of those properties. 

\begin{table*}
  \centering
  \caption{Table for isolated galaxies image simulation setups. 
  }
  \begin{threeparttable}
  \label{tab:iso_setup}
  \begin{tabular}{cccccc}
    \hline
    \noalign{\vskip 1mm}
    case & noise & PSF & galaxy & $10^3\mathrm{m} \: [3\sigma \: \mathrm{error}]$ & $10^4\mathrm{c} \: [3\sigma \: \mathrm{error}]$ \\
    \noalign{\vskip 1mm}
    \hline
    \noalign{\vskip 1mm}
    1 & fixed    & fixed Moffat      & Sersic & $0.5 \pm 1.8$ & $-0.08 \pm 0.35$ \\
    2 & variable & fixed Moffat      & Sersic & $0.8 \pm 1.8$ & $0.14 \pm 0.35$ \\
    3 & variable & variable Moffat   & Sersic & $0.8 \pm 1.8$ & $0.08 \pm 0.36$ \\
    4 & variable & variable Moffat   & Bulge+Disk & $0.3 \pm 1.9$ & $0.08 \pm 0.39$ \\
    \noalign{\vskip 1mm}
  \end{tabular}
  \end{threeparttable}
\end{table*}

Case 1 represents an idealized scenario with fixed noise and PSF, where galaxies are modeled as single-component Sersic profiles \citep{Sersic_profile}. The fixed PSF follows a Moffat profile with FWHM set to $0.7''$ for both wide- and deep-field images. The wide-field noise level is set to match the LSST $r$-band image noise level. The deep-field images are set to have $10\times$ less noise variance than the wide-field images. Case 2 introduces variable noise to assess whether \dacal matches the noise properties of the image correctly. 
We use the same setup as the previous case but independently scale the noise levels of the wide- and deep-field images by separate random factors drawn uniformly between 0.9 and 1.1 for each image. This introduces realistic mismatches in noise variance between the two fields. Any bias in our results will indicate imperfect noise matching between wide- and deep-field images and reflect the sensitivity of \dacal to such mismatches in noise properties. 
Case 3 further extends the previous case by incorporating variations in the size and shape of the PSF. 
 This test ensures that the PSF match between the wide- and deep-field samples is handled correctly in \dacal. We draw the wide- and deep-field PSF FWHM from a uniform distribution between $0.5''$ and $0.9''$ independently for each image.  
 Additionally, a small random shear is applied to the PSFs, with each component drawn uniformly between -0.02 and 0.02 independently for each image. Case 4 incorporates a more complex galaxy morphology by using a two-component bulge+disk model, which more accurately represents real galaxy structures. The bulge component follows the de Vaucouleurs profile \citep{DeVaucouleurs_profile} and the disk component follows an exponential profile. We use the same setup as in case 3.  


\subsection{Blended galaxies}
\label{sec:blended_setup}
Real survey images have blending of multiple objects. To evaluate the performance of \dacal in the presence of blending effects, 
we also tested our estimator on blended galaxy image simulations. This allows us to test the shear-dependent detection without the effect of redshift-dependent shear, which we defer to future studies. 
These image simulations are generated using the open source package \texttt{descwl-shear-sims}\footnote{\url{https://github.com/LSSTDESC/descwl-shear-sims}} \citep{metaDet_LSST2023}, which includes survey-specific parameters, including the image noise properties and PSFs of specific filters. 
The input galaxy catalog is drawn from another open-source package \texttt{WeakLensingDeblending}\footnote{\url{https://github.com/LSSTDESC/WeakLensingDeblending}}. The simulated images provide calibrated exposures with background subtraction and estimated noise variance. The input galaxy catalog includes model with bulge, disk, and active galactic nucleus (AGN) components. The bulge and disk components of the galaxy feature varying flux ratios, and the isophotes of these simulated galaxies are not strictly elliptical. Galaxy positions and orientations are randomly distributed across the image without accounting for spatial clustering effects.

Each simulated coadded image contains about $10^5$ input galaxies, covering $0.12 \:\mathrm{deg}^2$, corresponding to a number density of 230 galaxies per square arcmin. However, not all of these galaxies are detectable at the observational depths of LSST. We focus exclusively on the $r$-band for this analysis. 
The noise variance level of the wide-field images is set to match the expectations for a coadded image from 10 years of LSST observations. 

We applied the \texttt{descwl-shear-sims}-based pipeline (used for the blended simulations) to generate isolated-galaxy images as well. The resulting shear calibration biases were consistent with those obtained from the isolated-galaxy image simulation pipeline. This agreement indicates that the differences we observe in performance between the isolated and blended cases are genuinely due to the presence of blending, rather than artifacts of the different simulation codes used. 


\section{Results}
\label{sec:result}


In this section, we present the main results of our analysis. 
We first evaluate the shear calibration accuracy for isolated galaxies (Section~\ref{sec:result:isolated}) and extend the analysis to blended galaxies to assess the impact of shear-dependent detection biases (Section~\ref{sec:result:blended}). Finally, we examine the effect of sample variance in the deep-field images and its implications for the accuracy of the \dacal shear response (Section~\ref{sec:result:sample_variance}).

\subsection{\dacal performance for isolated galaxies}
\label{sec:result:isolated}


In this subsection, we quantify the accuracy of \dacal tested on isolated galaxy image simulations. We deliberately positioned the galaxy at the center of each stamp to avoid running any detection and selection process during the image processing. This is equivalent to setting the detection and selection weights in equation~\eqref{eq:nonlinear_ellip} to 1 for all galaxies.
The multiplicative and additive biases of \dacal in equation~\eqref{eq:mcbias} are measured as \citep{preciseSim_Pujol2019}

\begin{equation}
    \label{eq:mbias}
    m = \frac{\langle\hat{e}_\mathrm{dacal}^+ - \hat{e}_\mathrm{dacal}^-\rangle}{0.02\left\langle\widehat{R}_\mathrm{dacal}^+ - \widehat{R}_\mathrm{dacal}^-\right\rangle} - 1,
\end{equation}
and
\begin{equation}
    \label{eq:cbias}
    c = \frac{\langle\hat{e}_\mathrm{dacal}^+ + \hat{e}_\mathrm{dacal}^-\rangle}{\left\langle\widehat{R}_\mathrm{dacal}^+ - \widehat{R}_\mathrm{dacal}^-\right\rangle},
\end{equation}
where $0.02$ is the true shear value of constant shear used in the image simulations. The superscript `+' are quantities estimated from images distorted by positive shear and `-' are quantities estimated from negative shear.


\begin{table}
\centering
\begin{tabular}{lcc}
\hline
\noalign{\vskip 1mm}
Method & $10^3 m \: [3\sigma \: \mathrm{error}]$ & $10^4 c \: [3\sigma \: \mathrm{error}]$ \\
\noalign{\vskip 1mm}
\hline
\noalign{\vskip 1mm}
Wide-field only & $0.6 \pm 2.1$ & $1.0 \pm 3.9$ \\
Deep-field only & $-0.3 \pm 0.9$ & $-0.3 \pm 1.5$ \\
Deep + Wide (\dacal) & $0.9 \pm 1.3$ & $-1.4 \pm 2.4$ \\
\noalign{\vskip 1mm}
\hline
\end{tabular}
\caption{The multiplicative bias and additive bias for different calibration methods on blended image simulations with a fixed PSF size of 0.7 arcsec for both wide and deep fields. The top two rows correspond to applying \acal separately to the wide and deep fields, while the last row (``Deep $+$ Wide”) represents results from the \dacal method. The \dacal approach yields smaller uncertainties compared to applying \acal directly to the wide-field data. This comparison assumes equal survey areas for the wide and deep fields, so the distinction that is being tested has to do with the level of added noise in the images at fixed area.}
\label{tab:dacal_bias}
\end{table}

To evaluate the performance of \dacal, we conduct a series of controlled image simulations designed to quantity the accuracy of the shear estimation. 
In Table~\ref{tab:iso_setup}, we report the multiplicative and additive bias values for each case. In all cases, we find that the multiplicative bias is consistent with zero at the $3\sigma$ level and well within the LSST 10-year requirements. This indicates that our shear estimation remains unbiased under these isolated configurations. 

It is important to note that, in these tests, the number of simulated deep-field galaxies matches that of the wide-field galaxies, meaning the effects of sample variance for the deep field are not accounted for in this analysis. We also note that this test does not include detection or selection effects, which are known to introduce additional sources of bias in real analyses. Our results demonstrate that \dacal is an unbiased shear estimator for isolated galaxies, with multiplicative and additive biases consistent with expectations of second-order shear effects. 

\subsection{\dacal performance for blended galaxies}
\label{sec:result:blended}
In this subsection, we discuss the result of \dacal on blended galaxy image simulations. Unlike for the isolated galaxy image simulations, we must 
run the detection process to address detection bias due to blending. We use equations~\eqref{eq:mbias} and~\eqref{eq:cbias} to quantify the accuracy of \dacal in blended galaxy image simulations. 
 To generate each deep-field image, we randomly selected a new galaxy for every corresponding wide-field image, effectively neglecting sample variance in our analysis. This approach assumes that the wide- and deep-field data cover equal but independent areas of the sky. We defer the discussion of the effects of sample variance to Section~\ref{sec:result:sample_variance}.

Table~\ref{tab:dacal_bias} summarizes the results for different calibration strategies, including applying \acal separately to wide and deep fields, and using the combined method \dacal. Biases are reported in scaled units of $10^3m$ and $10^4c$ for clarity. 


The results demonstrate that in the unrealistic scenario that wide and deep fields have the same area, \acal applied to the deep field alone achieves a lower uncertainty due to its higher signal-to-noise ratio, as one would expect. The \dacal approach, which utilizes deep-field images for shear response calibration while preserving the statistical power of wide-field images, yields a multiplicate bias within the LSST requirement (gray shaded region). Furthermore, the \dacal method achieves smaller uncertainties compared to the standard implementation \acal on wide-field data alone. This improvement can be understood by considering the noise properties of each method: \acal applied to the wide-field dataset only includes two instances of the wide-field noise ($\sigma_w^2$), while \dacal incorporates one instance of the wide-field noise and two instances of the deep-field noise ($\sigma_d^2$). 
Given this ratio of noise contributions, we expect a statistical gain in the \dacal method given by $\sqrt{2\sigma_w^2 / (\sigma_w^2+2\sigma_d^2)} = \sqrt{2/(1 + 2/10)} \approx1.30$. These results highlight the effectiveness of \dacal in constraining the shear more precisely than \acal; both methods are consistent with being unbiased within our uncertainties. 

To further quantify the benefits of \dacal in the context of blended galaxy image simulations, we evaluate the effective number density of sources ($n_\mathrm{eff}$) following the approach of \citet{anacal_li}. The $n_\mathrm{eff}$ from blended galaxy image simulations is computed as
\begin{equation}
    \label{eq:neff}
    n_\mathrm{eff} = \left(\frac{0.26}{\sigma_\gamma}\right)^2 [\mathrm{arcmin}^{-2}],
\end{equation}
where $\sigma_\gamma$ is the uncertainty in the shear estimation. This approach allows us to calculate the number of galaxies with an intrinsic shape noise of 0.26 required to achieve $\sigma_\gamma$. Unlike the method used in \citet{WLsurvey_neffective_Chang2013}, which measures the effective galaxy number density by counting galaxies with specific weights, equation~\eqref{eq:neff} accounts for the correlation between the shape measurements of neighboring detections. Table~\ref{tab:neff} presents the effective number densities for different methods. When \acal is applied separately to the wide-field data we obtain $n_\mathrm{eff}=17$ arcmin$^{-2}$, while the deep-field data alone yields $n_\mathrm{eff}=41$ arcmin$^{-2}$, reflecting the higher signal-to-noise ratio of deep-field images. The \dacal method results in an intermediate effective number density of $n_\mathrm{eff}=30$ arcmin$^{-2}$. This improvement over the wide-field analysis demonstrates that \dacal retains a significant fraction of the statistical power of deep-field data, while avoiding the increased noise variance introduced in standard \acal applications to wide-field images. These results confirm that \dacal effectively mitigates the noise penalty associated with the latest implementation of standard \acal. The higher number density obtained with \dacal implies an improved precision in cosmic shear measurements, reducing statistical uncertainties in weak lensing analyses.

\begin{table}
    \centering
    \begin{tabular}{c|c}
        \hline
        Field & $n_\mathrm{eff}$  \\
        \hline
        Wide & 17 \\
        Deep & 41 \\
        Wide + Deep & 30\\
    \end{tabular}
    \caption{The effective number of galaxies per square arcminute used for weak lensing analyses, or $n_\mathrm{eff}$, derived using equation~\eqref{eq:neff} for the 10-year data from LSST. For all fields we use $r$-band images only. The first two rows are $n_\mathrm{eff}$ from using \acal on each respective fields. The last row is $n_\mathrm{eff}$ using the \dacal estimator.}
    \label{tab:neff}
\end{table}

\begin{figure}
    \centering
    \includegraphics[width=0.8\linewidth]{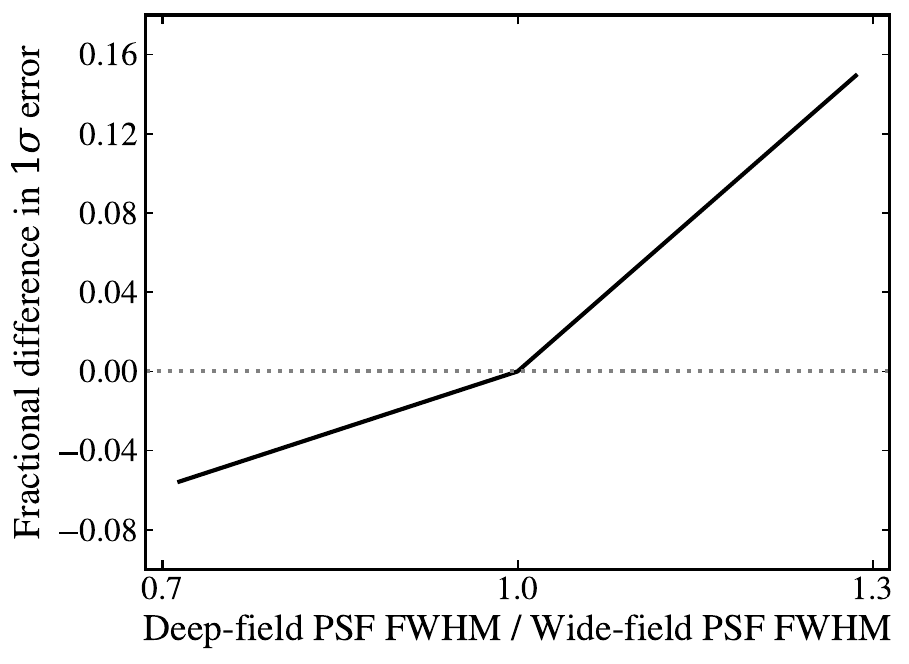}
    \caption{Fractional difference in the $1\sigma$ error on the inferred shear as a function of the ratio of the deep-field PSF size to the wide-field PSF size. The fractional difference is computed relative to the $1\sigma$ error when the deep-field and wide-field PSF sizes are equal (PSF FWHM ratio $= 1.0$). An increase in the deep-field PSF size relative to the wide-field PSF size increases the uncertainties on shear estimation. 
    }
    \label{fig:varying_psf}
\end{figure}

In a realistic survey, the deep field can have larger or smaller PSFs than the wide-field, depending on the observing conditions. We test the effect of PSF size variation on multiplicative and additive biases. For this setup, we simulate blended galaxy image simulations with the wide-field PSF FWHM fixed at $0.7''$, while the deep-field PSF FWHM is varied relative to the wide-field PSF FWHM. We find that the amplitudes of the multiplicative and additive biases remain unchanged, with only the uncertainties on these biases being affected.

Figure~\ref{fig:varying_psf} shows the fractional difference in the $1\sigma$ error in the inferred shear as a function of the ratio of the deep-field PSF size to the wide-field PSF size. The fractional differences are computed relative to the $1\sigma$ error when the deep-field and wide-field PSFs have the same size, such that it is zero in the plot. Our result indicates that a larger deep-field PSF size results in a larger uncertainty in the shear estimation. The upward trend can be understood using equation~\eqref{eq:nu_wide}: since the uncertainty in biases is dominated by the shape measurement uncertainties, and shapes are measured using the wide-field images, deconvolving the deep-field noise realization with a larger PSF in real space effectively corresponds to dividing the deep-field noise realizations by a smaller PSF in Fourier space. This process amplifies small-scale noise and thereby increases the uncertainty in the shape measurements. 
This result indicates that \dacal remains effective within the various observational conditions of a realistic survey. 

\subsection{Sample Variance}
\label{sec:result:sample_variance}
\begin{figure}
    \centering
    \includegraphics[width=\linewidth]{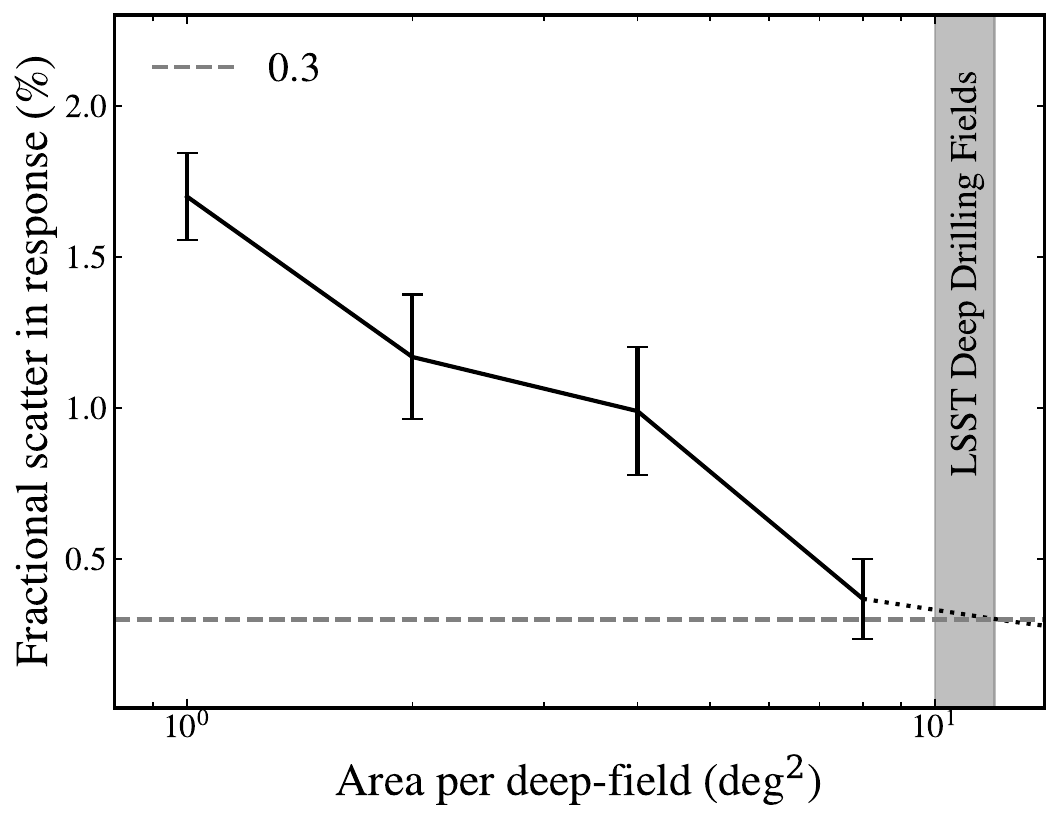}
    \caption{Fractional scatter in the \dacal response as a function of the area of a single deep-field patch, assuming a total of five independent  LSST Deep Drilling Fields, each with that area. 
    The dashed line corresponds to the requirements on the 10-year LSST shear calibration. The solid gray band shows the estimated area of LSST Deep Drilling Fields. The dotted line represents the extrapolations to the LSST Deep Drilling Fields area assuming the sample variance scales as $1 / \sqrt{\mathrm{area}}$.}
    \label{fig:samplev}
\end{figure}

In this subsection, we address the sample variance that was ignored in the previous analysis, which assumed equal areas for the wide and deep fields. LSST Deep Drilling Fields will cover $\sim$40 $\deg^2$ in five disjoint fields, each field having approximately $\approx 8$ $\deg^2$ of area \citep{Jones_2021_technote}. 
Due to the significantly smaller sky coverage of deep-field observations compared to wide-field surveys, the properties of deep-field galaxies (e.g., sizes, shapes) are subject to sample variance induced by large-scale structure. Since the \dacal response is computed using deep-field images and the properties of deep-field galaxies, this results in a larger error on the multiplicative bias. We characterize this effect by the fractional scatter in the \dacal response $\langle R_\mathrm{dacal}\rangle$. This fractional scatter can be directly compared with the precision required in the multiplicative bias, which is $|m| < 3\times10^{-3}$ for the final 10-year dataset of LSST \citep{LSSTDESC_2018arxiv}.

To estimate the effect of sample variance on the \dacal response, we used the OpenUniverse2024 \citep{OpenUniverse2024} catalog. The OpenUniverse2024 simulation was created to enable the production of matched simulated imaging data for multiple surveys observing a common simulated sky. 
All galaxies 
are constructed using a composite morphological model that includes distinct bulge, disk, and star-forming regions of the galaxies following a self-consistent star formation history, each with its own spectral energy distribution (SED). We refer readers to \citealt{OpenUniverse2024} for a comprehensive review of the simulation. The simulation provides a set of galaxy properties in a $70$~$\deg^2$ area, including object shapes and sizes that correlate with large-scale structure.

We follow the blended galaxy image simulation setup in Sec.~\ref{sec:blended_setup} and simulate deep-field images assuming the deep-field data have $10 \times$ more exposure time than the wide-field data, and we fix the PSF FWHM to $0.7''$ for both wide- and deep-field images. We also follow the strategy for LSST Deep Drilling Fields and \citealt{deep_metacal} and compute the scatter in \dacal response for a variety of deep-field areas, ranging from $1$ $\deg^2$ to $8$ $\deg^2$. Five disjoint fields with each area will be used to compute the scatter in the \dacal response. Figure~\ref{fig:samplev} shows the result of this computation. We find that the fractional scatter decreases as $\sim 1 / \sqrt{\mathrm{area}}$ as one might expect. Compared to \citet{deep_metacal}, we observe that the fractional scatter in our setup is about $1\%$ higher. The OpenUniverse2024 Simulation incorporates a more complex galaxy model than the catalogs used in \citet{deep_metacal}. Although the OpenUniverse2024 catalog has a higher galaxy number density, this difference in number density does not account for the observed increase in response sample variance scatter. 
For areas comparable to the LSST Deep Drilling Fields, we find that the uncertainty in the \dacal shear response due to sample variance is small enough that the overall shear calibration can still meet the LSST requirements. 
Our current results suggest that for a broad range of assumptions about the LSST Deep Drilling Fields, \dacal remains an effective technique. 

Further reduction in the sample variance within the \dacal response may be achievable by matching the properties of deep-field galaxies to those of wide-field galaxies. We defer this work to future studies.

\section{Conclusions}
\label{sec:conclusion}
In this work, we introduced \dacal, a novel extension of the \acal framework that utilizes deep-field observations to compute shear responses while preserving the statistical power of wide-field images. By leveraging the higher signal-to-noise ratio of detections in deep-field images, \dacal reduces the need to inject additional noise into wide-field survey data, improving the precision of shear measurements while maintaining calibration accuracy. 

We validated \dacal on both isolated and blended galaxy image simulations. Our results show that:
\begin{enumerate}
    \item \dacal achieves the required multiplicative bias accuracy of $|m| < 3\times10^{-3}$ (99.7\% confidence level), satisfying the shear calibration requirements for the 10-year LSST.
    \item Compared to the standard \acal applied on wide-field images, \dacal improves the effective galaxy number density from 17 to 30 galaxies per arcmin$^2$, a $76\%$ increase, for $r$-band coadded images with 10-year LSST depth.
    \item The uncertainty on the multiplicative bias is reduced by approximately $40\%$, demonstrating both improved accuracy and precision.
\end{enumerate}

To assess the impact of sample variance on deep-field images, we analyzed the expected variation in \dacal response under realistic deep-field survey strategies. Using the OpenUniverse2024 catalog, we estimated the fractional scatter in \dacal response and found that for deep-field areas comparable to the LSST Deep Drilling Fields, the sample variance contribution remains within the systematic error budget set for the weak lensing shear calibration of LSST.

Future work will explore additional optimizations, such as reweighing deep-field galaxies to match their properties with wide-field galaxies to further reduce sample variance effects in \dacal. Additionally, future work will also focus on extending this methodology to incorporate tomographic shear estimation. We expect that \dacal can be generalized to support tomographic analyses by computing redshift-dependent shear responses. 
Our findings suggest that \dacal represents a promising and computationally tractable shear calibration method for LSST and future weak lensing surveys, enabling more precise and robust cosmological constraints from cosmic shear.

\section*{Acknowledgements}
This work was performed using the Vera cluster at the McWilliams Center for Cosmology at Carnegie Mellon University, operated by the Pittsburgh Supercomputing Center facility. We thank the OpenUniverse2024 team for making their simulated galaxy catalogs publicly available.

Xiangchong Li is an employee of Brookhaven Science Associates, LLC under
Contract No. DE-SC0012704 with the U.S. Department of Energy. AP and RM were supported by the Department of Energy Cosmic Frontier program, grant DE--SC0010118, and by a grant from the Simons Foundation (Simons
Investigator in Astrophysics, Award ID 620789). Argonne National Laboratory’s work was supported under the U.S.
Department of Energy contract DE-AC02-06CH11357. 
This material is based upon work supported by the U.S. Department of Energy, Office of Science, Office of Workforce Development for Teachers and Scientists, Office of Science Graduate Student Research (SCGSR) program. The SCGSR program is administered by the Oak Ridge Institute for Science and Education for the DOE under contract number DE-SC0014664.

We thank the maintainers of numpy \citep{numpy}, \textsc{SciPy} \citep{2020SciPy-NMeth}, Matplotlib \citep{Matplotlib}, and \textsc{GalSim} \citep{galsim} for their excellent open-source software.

\section*{Data Availability}
No new data was generated for this work. The code used for this paper is publicly available on Github: \url{https://github.com/andyyPark/deep_anacal/tree/main}. The code used to analyze blended galaxy image simulations is \url{https://github.com/mr-superonion/xlens/tree/v0.2.2} and the code for shear estimation is available from \url{https://github.com/mr-superonion/AnaCal}.



\bibliographystyle{mnras}
\bibliography{main} 








\bsp	
\label{lastpage}
\end{document}